\documentclass[twocolumn,aps,prl,footinbib,showpacs]{revtex4}

\usepackage[sort&compress]{natbib}
\usepackage{graphicx}
 \usepackage{textcomp}
  \usepackage{amsmath}
\usepackage{lineno}
\usepackage{upgreek}
 
\setcitestyle{square,comma}

\newcommand{\re}{$\rm{^{87}Rb}$~}
\newcommand{\ket}[1]{|#1\rangle}
\newcommand{\ud}{\,\mathrm{d}}
\newcommand{\Z}{\mathcal{Z}}

\begin{document}

\title{Dynamics of a tunable superfluid junction}

\author{L.~J.~LeBlanc}
\author{A.~B.~Bardon}
\author{J.~McKeever}
\author{M.~H.~T.~Extavour}
\author{D.~Jervis}
 \author{J.~H.~Thywissen}
\affiliation{Department of Physics, University of Toronto, 60 St.~George, Toronto ON, Canada, M5S 1A7}
\author{F. Piazza}
\author{A. Smerzi}
\affiliation{INO-CNR, BEC Center, and Dipartimento di Fisica, Via Sommarive 14, 38123 Povo, Trento, Italy}

\date{{\today} }

\begin{abstract}

We study the population dynamics of a Bose-Einstein condensate in a double-well potential throughout the crossover from Josephson dynamics to hydrodynamics. At barriers higher than the chemical potential, we observe slow oscillations well described by a Josephson model. 
In the limit of low barriers, the fundamental frequency agrees with a simple hydrodynamic model, but we also observe a second, higher frequency.
A full numerical simulation of the Gross-Pitaevskii equation giving the frequencies and amplitudes of the observed modes between these two limits is compared to the data and is used to understand the origin of the higher mode.  Implications for 
trapped matter-wave interferometers are discussed.
\end{abstract}
\maketitle

Quantum mechanical transport is a consequence of spatial variations in
phase. Superfluids behave like perfect inviscid irrotational fluids,
whose velocity is the gradient of a local phase, so long as the
confining potential is smooth on the scale of the healing length.
Where the density is small, as it is near surfaces, quantum kinetic
terms must be added to the classical hydrodynamic equations. Macroscopic quantum coherence phenomena, such
as Josephson effects, emerge when superfluids are weakly linked across
such a barrier region.

Josephson effects have been demonstrated with superconductors \cite{AndersonPRL1963}, liquid helium \cite{PereverzevNature1997,BackhausScience1997}, and ultracold gases in both double-well \cite{AlbiezPRL2005,LevyNature2007} and multiple-well optical trapping potentials  \cite{LatticeMerge}. The canonical description of these experiments employs a two-mode model \cite{JosephsonPL1962, TMMMerge, TrombettoniPRA2003},  in which a sinusoidal current-phase relationship emerges.  
Hydrodynamics has also been studied in both liquids and ultracold gases \cite{JinPRL1996}.  The relative diluteness of gases makes a satisfying \emph{ab initio} description possible \cite{StringariZaremba}.  

In this Letter, we study the transport of a Bose-Einstein condensate (BEC) between two wells separated by a tunable barrier and observe the crossover from hydrodynamic to Josephson transport.  As the barrier height $V_\mathrm{b}$ is adjusted from below to above the BEC chemical potential, $\mu$, the density in the link region decreases until it classically vanishes when $V_\mathrm{b} = \mu$.  The healing length in the link region, $\xi$, increases with $V_\mathrm{b}$ and dictates the nature of transport through this region.  Oscillatory dynamics  spanning three octaves are observed as we smoothly tune $\xi$ from 0.3$d$ to 2$d$, where $d$ is the separation between the wells.

Examination of the dynamics of an elongated BEC in a double well is timely. Recent experiments have created squeezed and entangled states by adiabatically splitting a BEC \cite{JoPRL2007,JoFluctPRL2007, EsteveNature2008}.  The degree of squeezing inferred in the elongated case \cite{JoPRL2007,JoFluctPRL2007} seems to exceed what would be expected in thermal equilibrium \cite{EsteveNature2008}, raising the possibility that out-of-equilibrium dynamics may be important.  
With much remaining to be explored in these systems, this work represents the first study of the dynamics in the crossover regime.

 \begin{figure}[bt]
\begin{center}
\includegraphics[height = 3.6cm,clip=true,trim=0cm 0.0cm 0cm 0cm]{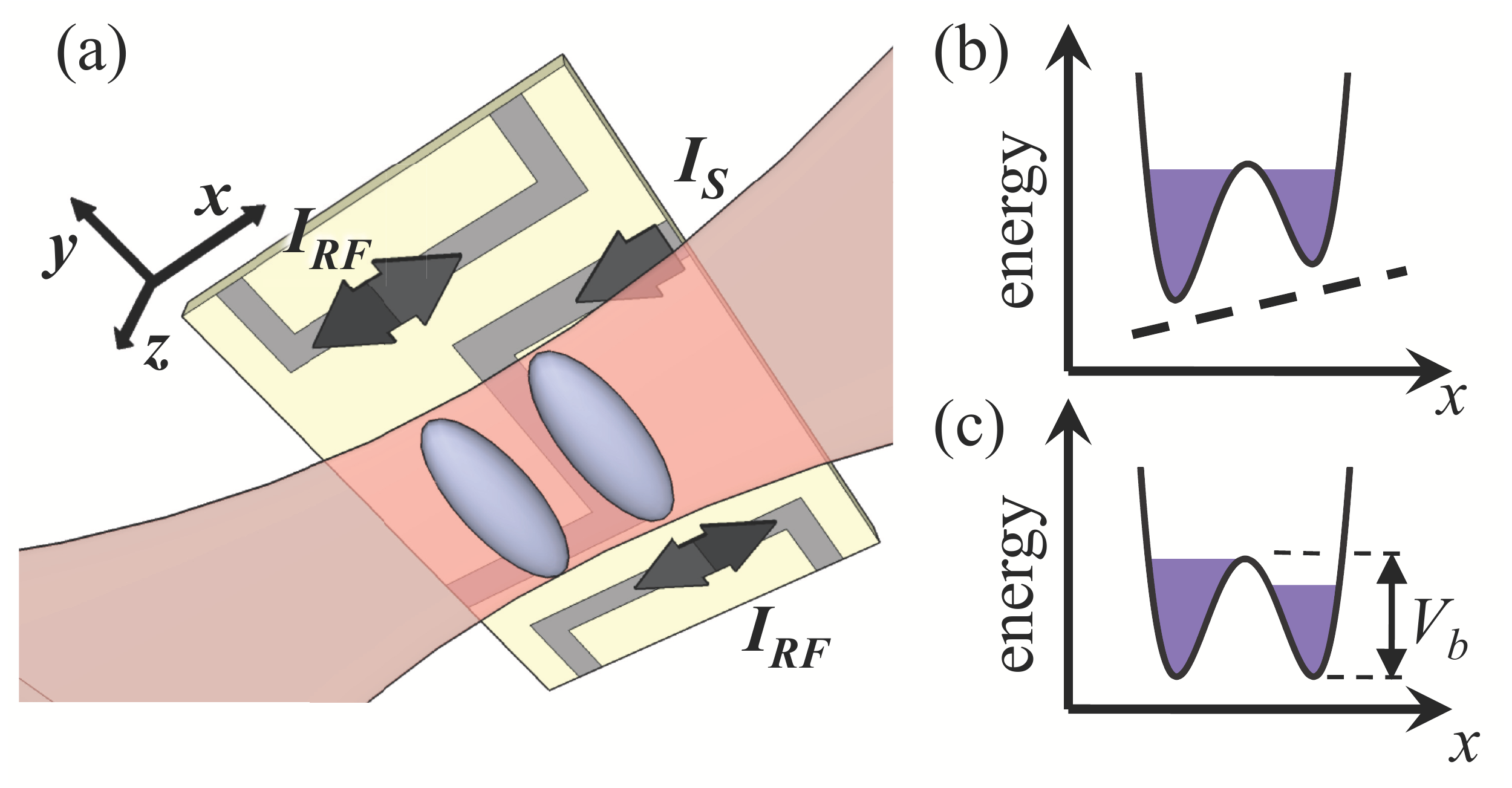}
\caption{(a)  Schematic of atom chip double-well trap. Central `Z'
wire \cite{ReichelAPB2002} carries static trapping current, $I_\mathrm{S} = 2$~A, which, with uniform external fields $\mathbf{B}_\mathrm{ext} = \langle2.2,0.11,0\rangle$ mT, results in an Ioffe-Pritchard style trap with harmonic trapping frequencies $(\omega_{x_0,z_0}, \omega_{y_0}) = 2 \pi \times (1300,10)$ Hz.  Side wires are 1.58 mm from trap center and carry RF currents with amplitude $I_\mathrm{RF}$.  This RF current produces a $z$-polarized field at the trap location with amplitude $B_\mathrm{RF} = 23.6 \pm 0.6~ \upmu$T (peak Rabi frequency $\Omega= 2 \pi \times (82\pm 2$ kHz)). A levitation beam (pink) is positioned to provide a force cancelling gravity ($z$-direction) while compressing the sample along $y$.  Atoms are trapped 190 $\upmu$m from the chip surface.
(b) A schematic one-dimensional cut at $t = -0.5$~ms through trapping potential along $x$ (solid line)  in the presence of linear bias (dashed line) and (c) balanced potential at $t=0$, with $\Z_0 \ne 0$. }
\label{fig:setup}
\end{center}
\end{figure}

Our experiment begins as  \re atoms in the $\ket{F = 2, m_F = 2}$ ground state are trapped on an atom chip and evaporatively cooled in a static magnetic potential $\mathbf{B}_\mathrm{S}(\mathbf{r})$, as described elsewhere \cite{AubinNatPhys2006}.  To prevent gravitational sag and to compress the trap in the weak direction (with characteristic trap frequency $\omega_y = 2 \pi \times 95$~Hz), we add an attractive optical potential with a 1064 nm beam.   We dress the static potential with an oscillating radio-frequency (RF) magnetic field \cite{ColombeEPL2004, LesanovskyPRA2006} radiating from two parallel wires on the atom chip (Fig.~\ref{fig:setup}(a)).    
In the rotating-wave approximation (RWA), the adiabatic potential created by the combination of the static chip trap, the RF dressing, and the optical force is 
\begin{align}
U(\mathbf{r}) = m'_F\mathrm{sgn}(g_F) \hbar \sqrt{\delta(\mathbf{r})^2 + \Omega_{\perp}^2(\mathbf{r})} + \tfrac{1}{2} m \omega^2_y y^2,
\label{eq:Ueff}
\end{align}
where $m'_F = 2$ is the effective magnetic quantum number, 
$\delta(\mathbf{r}) = \omega_\mathrm{RF} - \left|\mu_\mathrm{B} g_F B_\mathrm{S}(\mathbf{r})/\hbar\right|$ is the detuning, 
$\Omega_\perp(\mathbf{r}) = |\mu_\mathrm{B} g_F B_\mathrm{RF,\perp}(\mathbf{r})/2 \hbar|$  is the RF Rabi frequency, 
$B_\mathrm{RF,\perp}(\mathbf{r}) = |\mathbf{B}_\mathrm{S}(\mathbf{r}) \times \mathbf{B}_\mathrm{RF}(\mathbf{r})|/|\mathbf{B}_\mathrm{S}(\mathbf{r}) |$ is the amplitude of the RF field locally perpendicular to $\mathbf{B}_\mathrm{S}(\mathbf{r})$, 
$\mu_\mathrm{B}$ is the Bohr magneton, $g_F$ is the Land\'e g-factor, $\hbar$ is the reduced Planck's constant and $m$ is the atomic mass.  By assuming the individual wells are harmonic near each minimum, calculations show that $\omega_{z}= 2 \pi \times 425$~Hz, and $\omega_x$ varies from $2\pi \times 350$~Hz to $2 \pi \times 770$~Hz as we tune from low to high barriers.  For comparison between theory and experiment, we account for small corrections to Eq.~(\ref{eq:Ueff}) beyond the RWA \cite{HofferberthPRA2007,Supplementary}.

%The dressing field is turned on at a frequency $\omega_{RF}$, below the trap-bottom resonance of the static magnetic trap. Subsequent evaporative cooling produces a  BEC with no discernible thermal fraction.   In 20 ms, we adiabatically increase $\omega_{RF}$ from its initial value of $2\pi \times 765$~kHz to a new value characterized by $\delta_0 \equiv  \delta(\mathbf{r} = \mathbf{0})$.  As the detuning increases and the barrier $V_{\rm b}$ rises, the dressed state potential splits  along the $x$-direction into two elongated traps \cite{SchummNatPhys2005}.  

After turning on the dressing field at a frequency $\omega_{RF} = 2\pi \times 765$~kHz, where the trap is a single well, we evaporatively cool to produce a  BEC with no discernible thermal fraction.   In 20 ms, we adiabatically increase $\omega_{RF}$ to a new value characterized by $\delta_0 \equiv  \delta(\mathbf{r} = \mathbf{0})$,  such that the barrier $V_{\rm b}$ rises and the dressed state potential splits  along the $x$-direction into two elongated traps \cite{SchummNatPhys2005}.

Using a second 1064 nm beam weakly focussed off-center in $x$, an approximately linear potential is added across the double-well junction to bias the population towards one well (Fig.~\ref{fig:setup}(b)).  By applying the bias beam before and during the splitting process, we prepare systems of atoms with a population imbalance $\mathcal{Z} \equiv (N_\mathrm{R} - N_\mathrm{L})/(N_\mathrm{R} + N_\mathrm{L})$, where $N_\mathrm{R}$ ($N_\mathrm{L}$) is the number of atoms in the right (left) well.  The range of initial population imbalances $\Z_0 = \mathcal{Z}(t = 0)$ we use is 0.05 to 0.10, small enough to avoid self-trapping \cite{AlbiezPRL2005}.  To initiate the dynamics, the power of the bias beam is ramped off in 0.5 ms (faster than the population dynamics) and the out-of-equilibrium system is allowed to evolve for a variable time $t$ in the symmetric double-well (Fig.~\ref{fig:setup}(c)).

\begin{figure}[bt]
\begin{center}
\includegraphics[scale=0.350,clip=true,trim=0cm 0.0cm 0cm 0cm]{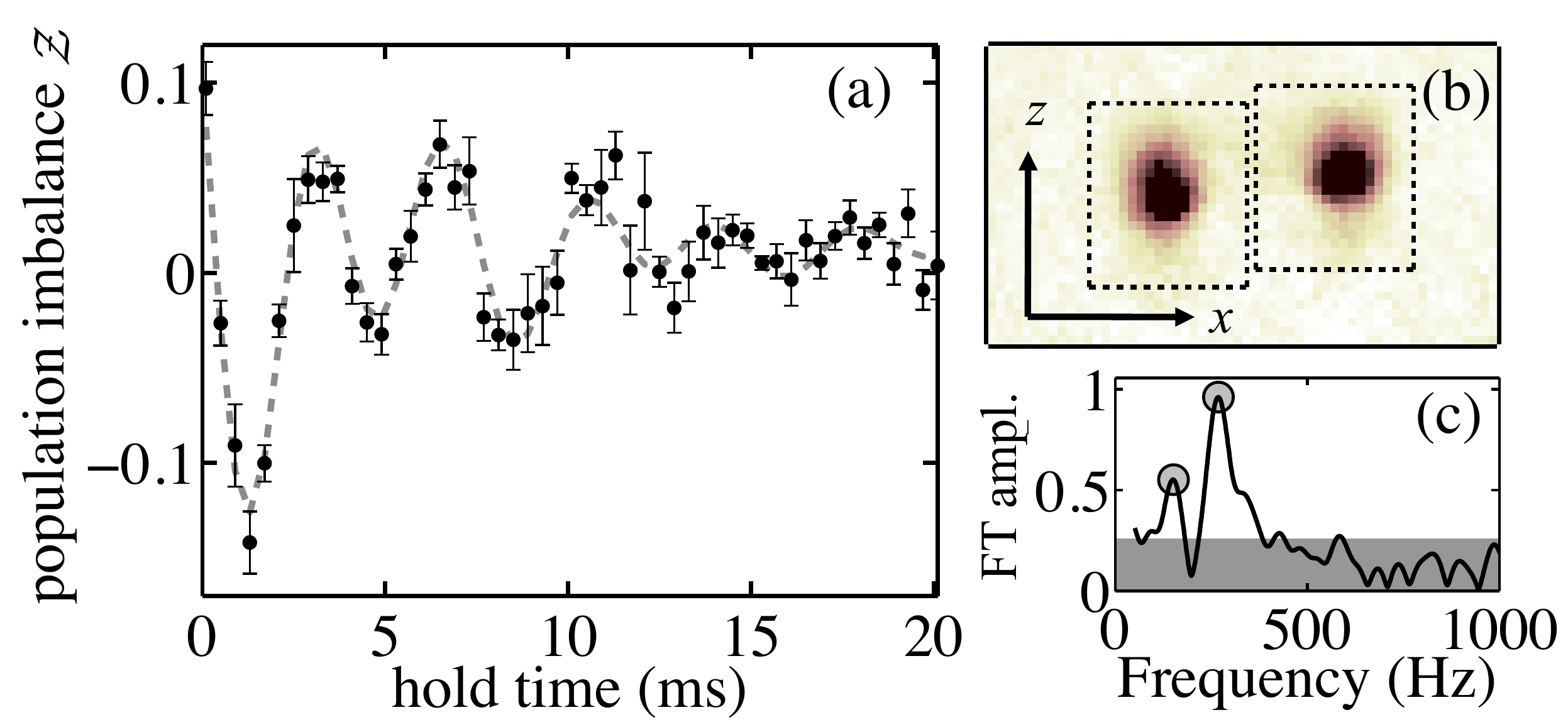}
\caption{(a) Population imbalance, $\mathcal{Z}$, vs.\ time for $\delta$ = $2\pi \times (0.1 \pm 0.5)$ kHz, $N = 5900 \pm 150$.  The dashed line is a decaying two-frequency sinusoidal fit to the data, using two fixed frequencies from the FT (lower inset).  Each point is the average of six repetitions of the experiment; error bars are statistical.  (b) Averaged absorption image after separation and 1.3 ms time of flight, with right and left measurement regions (dashed boxes) indicated.  (c)  FT amplitude spectrum of data  showing two distinct peaks at $268 \pm 6$ and $151 \pm 13$ Hz rising above the noise floor (grey).}\label{fig:timeseries}
\end{center}
\end{figure}

To measure the time-dependent population $\mathcal{Z}(t)$, we freeze dynamics by rapidly increasing both $B_\mathrm{RF}$ and $\omega_\mathrm{RF}$ to separate the wells by $70 ~\upmu$m, where $V_\mathrm{b}/\mu \sim 10^4$.  We release the clouds from the trap and perform standard absorption imaging along $y$ after 1.3 ms time-of-flight (Fig.~\ref{fig:timeseries}(b)).  Analysis of these images allows us to determine $N_\mathrm{R}$ and $N_\mathrm{L}$ to a precision of $\pm$50 atoms.
        
\setlength{\belowcaptionskip}{0pt}
\begin{figure}[bt]
\begin{center}
\includegraphics[scale=0.4,clip=true,trim=0.0cm 0.0cm 0cm 0cm]{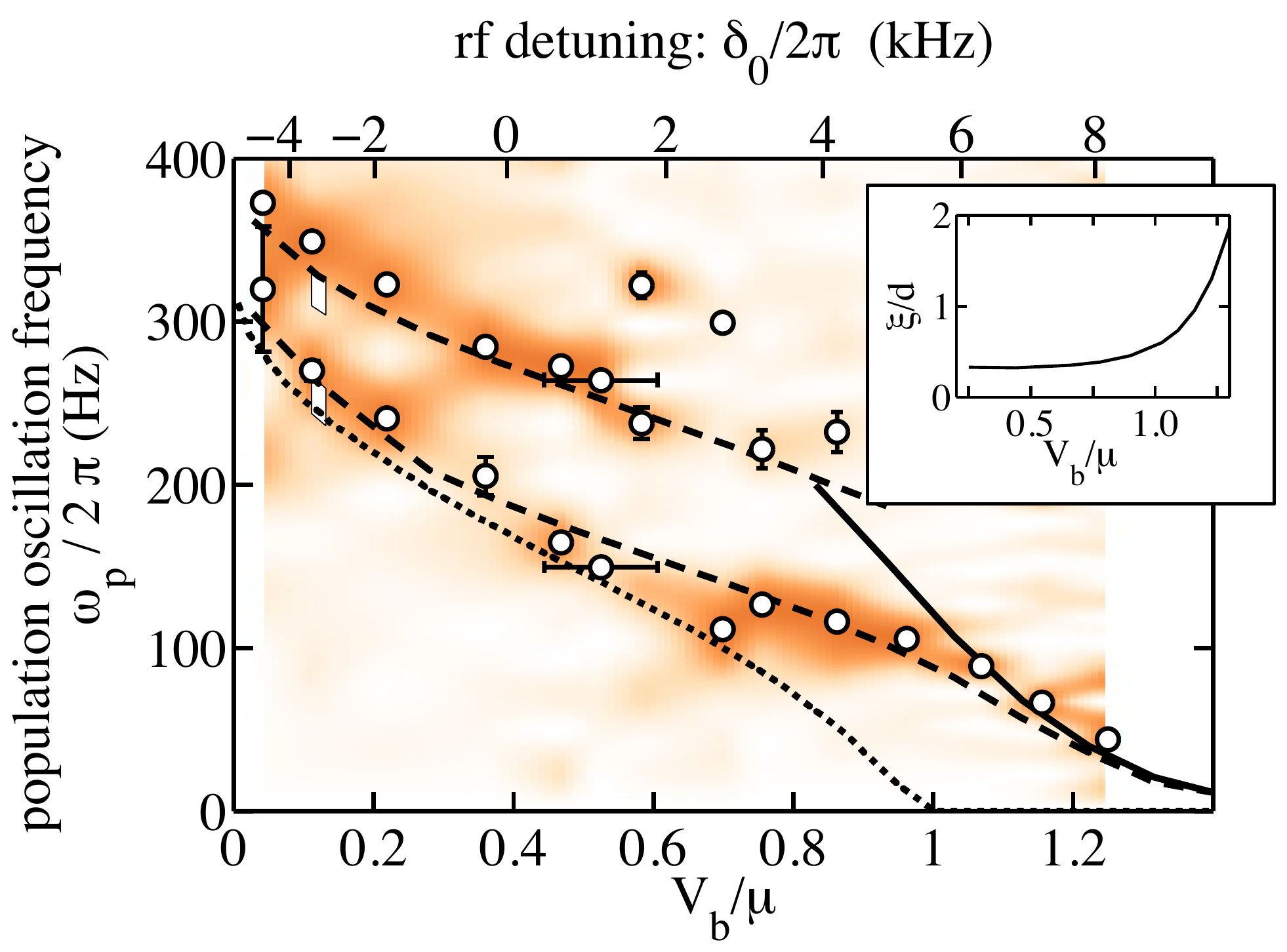}
\caption{Frequency components of population imbalance vs.\ RF detuning (measured) and barrier height to chemical potential ratio (calculated).  Experimental points (white circles) represent the two dominant Fourier components at each detuning; error bars represent uncertainty contributed by noise in the FT from a single time series, but do not include shot-to-shot
fluctuations.
The spectral weight is represented through the color map, which has been linearly smoothed between discrete values of $V_{\rm b}/\mu$ and darker colors indicate greater spectral weight.  All calculations use N = 8000 and $\Z_0 = 0.075$, and  a single-parameter fit of the data to the GPE curves shifts all experimental points by $\delta_\mathrm{shift} = 2\pi \times 5.1$~kHz \cite{Supplementary} to compensate for a systematic unknown in $B_\mathrm{S}(\mathbf{0})$. Statistical vertical error bars are shown, while a typical horizontal statistical error bar is shown at $V_\mathrm{b}/\mu \approx 0.5$.  
Dashed lines represent 3D GPE frequencies, the solid line the plasma oscillation frequency predicted by the Josephson model, $\omega_\mathrm{p}$, and the dotted line the hydrodynamic approximation, $\omega_{\rm HD}/2\pi$.
 White bars at $V_{\rm b}/\mu \sim 0.1$ indicate the bounds of the GPE simulation corresponding to the systematic plus statistical uncertainty in atom number.   Inset:  ratio of healing length, $\xi$, to interwell distance, $d$, as a function of $V_{\rm b}/\mu$.  $\xi$ is calculated at the center of the barrier. }\label{fig:freqdelta}
\end{center}
\end{figure}

Upon release of the potential bias, we find that the population $\mathcal{Z}(t)$
oscillates about  $\mathcal{Z} = 0$ (Fig.~\ref{fig:timeseries}(a)) \cite{footnoteZ}.   To analyze the dynamics, we use a Fourier transform (FT) to find the dominant frequency components (Fig.~\ref{fig:timeseries}(c)).  We repeat this measurement at many values of $V_{\rm b}/\mu$, where $\mu$ is the Thomas-Fermi chemical potential, by varying $\delta_0$.   For the purposes of this analysis, we ignore the decay of this signal, the $1/e$ time constant of which is typically two oscillation periods. 

When the barrier is low, $\mathcal{Z}(t)$ consistently displays two dominant frequency components.
For higher barriers, the amplitude of the higher-frequency mode decreases until only a single frequency rises above the noise floor.  The white points in Fig.~\ref{fig:freqdelta} give these frequencies as a function of the experimental parameter $\delta_0$ and the calculated ratio of barrier height to chemical potential, $V_{\rm b}/\mu$.  The ensembles used in Fig.~\ref{fig:freqdelta} had total atom number $N = 6600 \pm 400 ~(\pm 1700)$, where the error bar is statistical (systematic).  
		
In the low- and high-barrier limits, simple models can be used to understand the dynamics.  For low barriers, the hydrodynamic equations of motion can be used to estimate the frequency
of population oscillation. Assuming a harmonic population response for some $\Z_0$, the response
frequency is
\begin{align}
 \omega_{\rm HD}^2 \approx -\frac{2}{mN\mathcal{Z}_0} \int_S \rho \, \hat{n} \cdot \vec{\nabla}(U+g\rho) dS,
\end{align}
where $\rho$ is the density of the condensate at $t=0$, $S$ is the surface in the $y$-$z$ plane bisecting the double well, and $\hat{n}$ is the vector normal to this surface.  Plotting $\omega_{\rm HD}$ in Fig.~\ref{fig:freqdelta} (dotted line), we find good agreement with the lower frequency mode at low barriers. Since tunnelling cannot contribute to hydrodynamic transport, $\omega_\mathrm{HD} \rightarrow 0$ as $V_\mathrm{b} \rightarrow \mu$. The breakdown in hydrodynamics also coincides with an increasing healing length, as shown in the inset of Fig.~\ref{fig:freqdelta}.  

In the opposite limit, when tunnelling dominates transport, a Josephson model \cite{TMMMerge} accurately predicts the frequency of the highest barrier points, 
\begin{align}
\omega_{\rm p}^2 = \frac{1}{\hbar^2}\Delta E \left(\Delta E + N\frac{\partial \mu_{\mathrm loc}}{\partial N} \right), 
\end{align}
where $\Delta E$ is the energy difference between the symmetric and antisymmetric ground states of the double-well potential, $N$ is total atom number,  and $\mu_\mathrm{loc}$ is the chemical potential on one side of the well \cite{Supplementary}.
 The agreement is surprisingly good even for $V_{\rm b}$ just above $\mu$, beyond which the frequency decreases exponentially. To our knowledge, this constitutes the first direct observation of tunneling transport of neutral atoms through a magnetic barrier, only inferred, for instance,  in Refs.~\cite{JoPRL2007, MaussangArxiv2010}.		
		
To explain the crossover behavior and the existence of the higher-frequency mode, we turn to numerical solutions of a time-dependent three-dimensional Gross-Pitaevskii equation (GPE) \cite{TMMMerge, GPE}, which should describe all mean-field dynamics at $T = 0$.  The slope and separation of the measured frequencies are well captured by the GPE, as shown in Fig.~\ref{fig:freqdelta}, though the decay of population imbalance is not reproduced by these simulations.

The structure and origin of the higher-lying dynamical mode can be studied within the simulations. If our trap were smoothly deformed to a spherical harmonic potential, the two observed modes would connect to odd-parity modes \cite{StringariZaremba}: the lower mode connects to the lowest $m=0$  mode  (coming from the $\ell = 1$  mode at spherical symmetry, where the quantum numbers $\ell$ and $m$ label the angular momentum of the excitation and its projection along the axis of symmetry, $y$, respectively), while the higher mode originates from the lowest $m = 2$ mode ($\ell = 3$ at spherical symmetry) \cite{footnoteTMM}. 

With insight from GPE simulations, the observation of a second dynamical mode, which was not seen in previous experimental work \cite{AlbiezPRL2005,LevyNature2007}, can be explained.  In a purely harmonic trap, a linear bias excites only a dipole mode  \cite{KohnPR1961}.
By breaking harmonicity along the splitting direction, $x$, the barrier allows the linear perturbation ($\ell = 1, m = 0$, where $x$ is the azimuthal axis) to excite multiple Bogoliubov modes  \cite{ZimmermannPRL2003}. Numerical studies show that two additional ingredients are required to excite the higher mode.  First, atom-atom interactions couple the $x$-excitation to the transverse ($y,z$) motion through the nonlinear term in the GPE.  Second, the anisotropy of the trap in the $y$-$z$ plane mixes the $m=0$ and  $m = 2$ modes such that each of the resulting modes drives population transfer between wells.

\begin{figure}[bt]
\begin{center}
\includegraphics[scale=0.4,clip=true,trim=0cm 0.0cm 0cm 0cm]{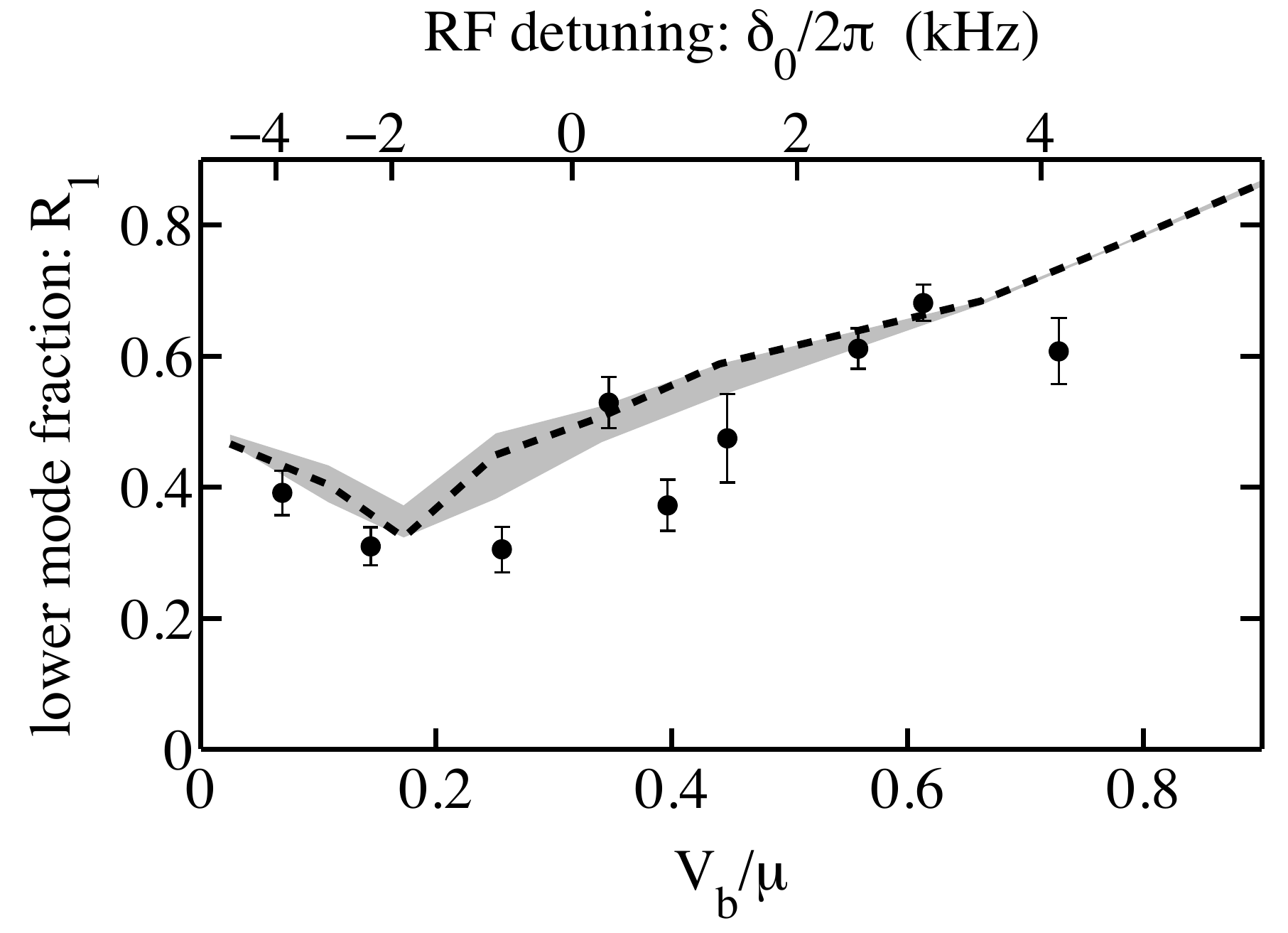}
\caption{Fraction of low-frequency mode in population dynamics. Dashed line shows the GPE simulation for 8000 atoms with initial imbalance $z(0) = 0.075$.  The grey shaded area represents the variation of the GPE calculations over the range of z(0) = 0.05 to 0.10.  The vertical error bars are statistical; the statistical uncertainty in $\delta$ is $2\pi \times 0.5$~kHz (not shown).  The GPE calculation gives $R_1 = 1$ when $V_{\rm b}/\mu \simeq 1.1$. } \label{fig:amplitude}
\end{center}
\end{figure}

Figure~\ref{fig:amplitude} shows the relative strength $R_1 = a_1/(a_1+a_2)$ of the lower frequency mode as a function of the barrier height. The amplitude $a_1$ ($a_2$) of the lower (higher) frequency mode is extracted from a decaying two-frequency sinusoidal fit.  The modes have comparable strength, even in the linear perturbation regime, when the barrier is below the chemical potential.   The small spread in the GPE amplitudes  shown by the grey band indicates that the higher mode is excited independently of the initial imbalance, and is not simply due to a high-amplitude nonlinearity.

The trend in $R_1$ reflects the shape of the trap.  When the barrier is raised from zero, the higher mode is at first more easily excited due to an increased anharmonicity along $x$ as the trap bottom becomes flatter. By further increasing the barrier, the higher-frequency mode disappears from the population oscillation spectrum due to the vanishing excitation of transverse modes.  As the wavefunctions in each individual well are increasingly localized to the effectively harmonic minima, the linear bias no longer excites  intrawell transverse motion. Furthermore, in the linear perturbation regime, the interwell Josephson plasma oscillation, like all Bogoliubov modes, cannot itself trigger any other collective mode.  

In conclusion, we have studied the quantum transport of a BEC in a double-well potential throughout the crossover from hydrodynamic to Josephson regimes.
Apart from fundamental interest, knowing and controlling the nature of
superfluid transport is crucial for technological applications of
weak-link based devices, such as double-slit interferometers \cite{JoPRL2007, SchummNatPhys2005, ShinPRL2005, GrossNature2010, BaumgartnerArxiv2010}. The adiabatic transformation of a BEC from a single- to a
double-well trapping potential has been discussed in recent
experimental works  \cite{JoPRL2007, MaussangArxiv2010,EsteveNature2008,WangPRL2005,JoPhasePRL2007} in the context of the Josephson model, valid
at high barriers \cite{LeggettPRL1998}.  
Our work demonstrates that for $V_{\rm b}< \mu$, the
lowest mode frequency will lie below that estimated by the Josephson model.
Furthermore, the higher-lying mode we observe approaches the lowest
collective mode as $\omega_y \ll \omega_z$ \cite{Supplementary} and may be important to
the dynamics of splitting in strongly anisotropic double wells
\cite{JoPRL2007, PolkovnikovNatPhys2008}.  Whether using splitting to prepare entangled
states \cite{EsteveNature2008}, or recombination \cite{JoPhasePRL2007} to perform closed-loop
interferometry \cite{WangPRL2005}, an improved understanding of double-well dynamics
provides a foundation for controlling mesoscopic superfluids.

\begin{acknowledgements}
We would like to thank T.~Schumm for early experimental work, A.~Griffin,  P.~Kr\"uger, D.~McKay, M.~Sprague, and E.~Zaremba for helpful discussions, and J. Chwede\'nczuk for help with numerical simulations of the GPE.  This work has been generously supported by CIfAR, CFI, CQIQC, and NSERC. 
\end{acknowledgements}

\bibliographystyle{h-physrev.bst}

\onecolumngrid

\pagebreak
\hspace{20pt}
\begin{center}
\large{\bf{Supplementary materials for ``Dynamics of a tunable superfluid junction''}}
\hspace{60pt}
\end{center}

\setcounter{figure}{0} 
\makeatletter 
\renewcommand{\thefigure}{S\@arabic\c@figure} 
\renewcommand{\thesection}{S\@Roman\c@section} 
\renewcommand{\thetable}{S\@arabic\c@table} 
\makeatother 

\setcounter{equation}{0} 
\makeatletter 
\renewcommand{\theequation}{S\@arabic\c@equation} 
\makeatother 

\hspace{100pt}
\twocolumngrid

\section{Josephson model}
We compare our experimental results to the Josephson model (JM) and its plasma frequency, $\omega_p$ (Fig.~3).  The JM employed here is based on the nonlinear two-mode ansatz used in \cite{TrombettoniPRA2003},
\begin{equation}
\Psi(\mathbf{r}, t) = \psi_R\phi_R(\mathbf{r};N_R(t)) +   \psi_L\phi_L(\mathbf{r};N_L(t)) 
\end{equation}
where  $\psi_{R,L} = \sqrt{N_{R,L}(t)} \exp(i\theta_{R,L}(t))$ and $\phi_{R,L} =  (\phi_+ \pm \phi_-)/\sqrt{2}$ is a real function localized
in the left (right) well, with $\phi_+$ ($\phi_-$) being the ground (first antisymmetric) state of the GPE along the splitting direction.
The linearized equation gives $\mathcal{Z}(t) = \mathcal{Z}(0) \cos(\omega_pt + \Delta \theta(0))$, where $\Delta \theta = \theta_R - \theta_L$ and the plasma frequency
\begin{equation}
\omega_p^2 = \frac{1}{\hbar^2}\Delta E \left(\Delta E + N \frac{\partial \mu_{loc}}{\partial N_{\rm L}}\right),
\end{equation}
where
\begin{equation}
\mu_{loc} = \int \ud \mathbf{r} \left[ \frac{\hbar^2}{2m} (\nabla \phi_{R,L} )^2 + U(\mathbf{r}) \phi_{R,L} ^2 + gN_{R,L} \phi^4_{R,L}  \right],
\end{equation}
and $\Delta E = E_- - E_+ = 2 (\mathcal{K} + N\chi)$ with
\begin{eqnarray}
E_{\pm} &=& \int \ud \mathbf{r} \left[ \frac{\hbar^2}{2m} (\nabla \phi_{\pm})^2 + U(\mathbf{r}) \phi_{\pm}^2 + \frac{1}{2}gN\phi^4_{\pm} \right],\\
\mathcal{K} &=& -\int \ud \mathbf{r} \left[ \frac{\hbar^2}{2m}  (\nabla \phi_{R})  (\nabla \phi_{L})+ \phi_{R}U(\mathbf{r}) \phi_{L} \right],\\
\chi &=& -\frac{g}{4} \int \ud \mathbf{r} \phi_R^3 \phi_L.
\end{eqnarray}

 The plasma frequency $\omega_p$ depends on the derivative of the single-well chemical potential $\mu_\mathrm{loc}$, and therefore takes into account the effect of transverse degrees of freedom on the effective nonlinearity determining the interaction energy. This provides an important correction to the plasma frequency, typically around 20\%.  Here $\Delta E$ is the energy splitting between the ground state and the lowest antisymmetric state along the splitting direction.  In our experiments, for example, $\mu_\mathrm{loc}/\hbar = 2\pi \times 1.8~$kHz and $\Delta E /\hbar= 2\pi \times 3.7$~Hz at $\delta_0 = 6.9$~kHz.

\section{Hydrodynamic model}

To determine the behaviour of the condensate in a double well in the hydrodynamic regime, we use the continuity equation and the equation of motion for the condensate in the hydrodynamic description, ignoring the quantum pressure term:
\begin{align}
&\frac{\partial \rho(\mathbf{r},t)}{\partial t} + \nabla\cdot\left[\mathbf{v_\mathrm{s}}(\mathbf{r},t) \rho(\mathbf{r},t)\right] = 0 \label{eq:continuityGPE}\\
&m\frac{\partial \mathbf{v_\mathrm{s}}(\mathbf{r},t)}{\partial t} + \nabla\left[ U(\mathbf{r}) + g \rho(\mathbf{r},t)  + \tfrac{1}{2} m\mathbf{v}_\mathrm{s}^2(\mathbf{r},t) \right] = 0
\label{eq:hydroGPE}
\end{align}
where $\rho(\mathbf{r},t)$ is the local density and $\mathbf{v_\mathrm{s}}$ is the superfluid velocity.  We assume harmonic motion of the population balance between the wells such that
\begin{equation}
\ddot{\mathcal{Z}} = -\omega_{\rm HD}^2 \mathcal{Z} \label{eq:sho}
\end{equation}
where $\omega_{\rm HD}$ is the hydrodynamic frequency that characterizes the system.  

The first time derivative of $\Z\equiv 2 N_\mathrm{R}/N$ is
\begin{align}
\dot{\Z}  =&~  \frac{2}{N}\int_{V_R}  \dot{\rho} \, \ud ^3 \mathbf{r}\nonumber \\
 =&~  - \int_{V_R} \!\!\!\nabla \cdot ( \rho \mathbf{v}_s ) \,\ud ^3 \mathbf{r}
 =  - \int_S \hat{n} \cdot  ( \rho \mathbf{v}_s  ) \,\ud S
\end{align}
where $V_R$ is the volume of the right well, $S$ is the area of the plane separating the two wells, and $\hat{n}$ is the unit normal vector for this plane. 

The second derivative of $\Z$ is then
\begin{equation}
\ddot{\mathcal{Z}} 
=  -\frac{2}{N} \int_S \hat{n} \cdot  ( \dot{\rho} \mathbf{v}_s + \rho \dot{\mathbf{v}_s}  ) \,\ud S.
\end{equation}
To evaluate the frequency, $\omega_{\rm HD}$, we assume that the system begins at rest, such that $\mathbf{v}_s(t = 0) = 0$.  The time derivative of $\mathbf{v}_s$ is given by the hydrodynamic equation of motion, Eq.~(\ref{eq:hydroGPE}), and 
\begin{equation}
\left. \ddot{\mathcal{Z}} \right|_{t=0} \mathbf{=}  \frac{2}{m N} \int_S \rho \, \hat{n} \cdot \vec{\nabla} \left(U(\mathbf{r}) + g \rho\right) \,\ud S.
\end{equation}
The geometry of this double well system is such that the normal vector $\hat{n} = \hat{x}$, and the only component of the gradient which contributes is the $x$-component.  Assuming some initial imbalance, $\Z_0$, the frequency with which the populations oscillate is given by
\begin{equation}
\omega^2_\mathrm{HD} = -\frac{\ddot{\mathcal{Z}}}{\mathcal{Z}} \mathbf{=} - \frac{2}{m N \mathcal{Z}_0}\int \!\! \int_S \rho  \,\, \frac{\partial}{\partial x} \left(U(\mathbf{r}) + g \rho \right)  \ud y \ud z. \label{eq:spring}
\end{equation}

We calculate this initial density profile in the trap, tilted by a linear bias $G x$, using the Thomas-Fermi approach:
\begin{equation}
\rho_\mathrm{TF}(\mathbf{r}) = \left(\mu - (U(\mathbf{r}) + Gx)\right)/g,
\end{equation}
The gradient term in the integrand of Eq.~(\ref{eq:spring}) is then simply $-G$.

The characteristic frequency is thus
\begin{equation}
\label{eq:omega}
\omega^2_\mathrm{HD} \approx  \frac{2 G}{m N \mathcal{Z}_0} \int \! \! \int_S \rho_{\rm TF} \, \ud y  \ud z,
\end{equation}
which indicates that the frequency can be found by simply evaluating the density at the surface between the two wells and integrating over the region by which the two halves are connected.  From this expression, we see that the $\omega^2_\mathrm{HD}$ decreases as the area connecting the wells decreases, and falls to zero when the barrier surpasses the chemical potential and the Thomas-Fermi density is strictly zero on the plane $S$.

The equation (\ref{eq:spring}), upon substituting $\rho_{\rm TF}$ with the Gross-Pitaevskii ground state density, is also valid when we include a quantum pressure term in Eq.~(\ref{eq:hydroGPE}). However, even with the quantum pressure, this model is not in exact agreement with the Gross-Pitaevskii equation, due to the non-harmonic component in the oscillation.
At high barriers, though anharmonicity is small, Eq.~(\ref{eq:spring}) is less accurate than the JM.

We have checked that the frequencies predicted by Eq.~(\ref{eq:omega}) are consistent with the dynamical simulations of Eqs.~(\ref{eq:continuityGPE}),~(\ref{eq:hydroGPE}), done using a test particle method [S2].

\section{Gross-Pitaevskii Equation}
We solve numerically the time-dependent equation
\begin{equation}
i\hbar\partial_t \Psi(\mathbf r) = -\frac{\hbar^2}{2m} \nabla^2\Psi(\mathbf r) + \left(U(\mathbf{r}) + g|\Psi(\mathbf r)|^2\right) \Psi(\mathbf r)
\end{equation}
where $\Psi(\mathbf r)$ is the complex condensate order parameter, $U(\mathbf r)$ is the double-well external trapping potential, and $g=4\pi\hbar^2 a_s/m$ with $a_s$ the \re s-wave scattering length. 
While all calculations were done using Eq.~(\ref{eq:Ueff}), an intuitive understanding of the potential emerges from the separable approximate form:
\begin{equation}
U_{sep}\simeq \tfrac{1}{2} m \omega_y^2 y^2+ \tfrac{1}{2} m\omega_z^2 z^2 + \alpha_2 x^2+ \alpha_4 x^4\ ,
\label{eqpot}
\end{equation}
where $\alpha_2 < 0$.  The trap exhibits an axial anisotropy, with elongation along the $y$-direction such that $\omega_z\sim 4 \omega_y$.

\section{Decay of population imbalance}
In comparing the time series measured in the experiment with those found from GPE calculations, we see similar multiple-frequency behaviour.  One striking difference is the presence of ``decay'' in the experimental data -- the fall off of the amplitude of the populations oscillations with time.  The characteristic time scale of the decay, $\tau$, is approximately equal to two oscillation periods over all values of $V_\mathrm{b}/\mu$.  We model this as an exponentially decaying envelope in our analysis, and include it in our fitting equation (Eq.~(\ref{eq:2freq})).  

\begin{figure}
\includegraphics[scale=0.43,clip=true,trim=0.0cm 0.0cm 0cm 0cm]{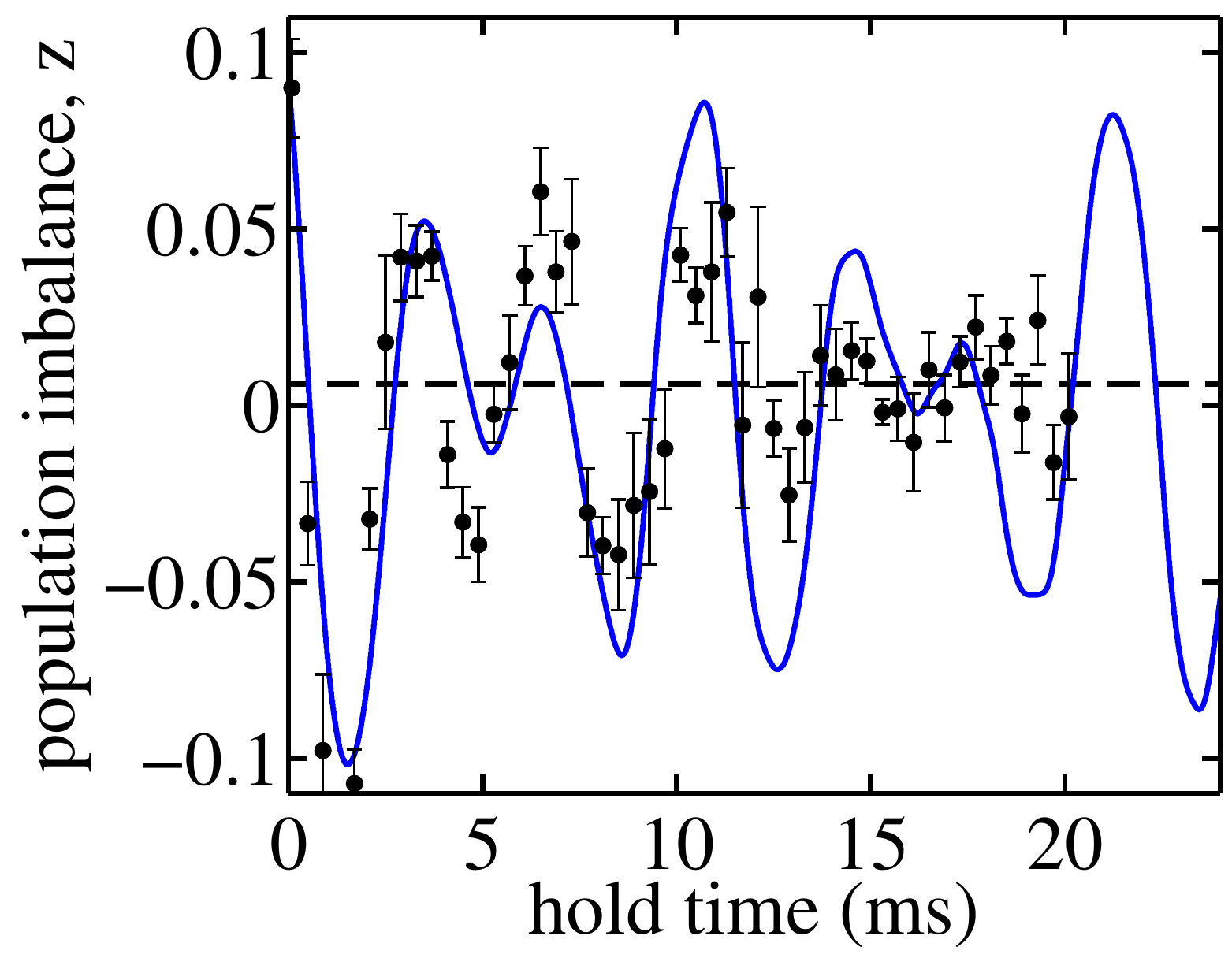}
\caption{Comparison of experimental and GPE time series for $\delta = -0.1$~kHz (GPE)) and  $\delta = -0.1 \pm 0.5$~kHz (experiment).  Experimental points are shown as black dots, and the fit to experimental data is shown as a black dashed line.  The GPE results are shown as a solid blue line.  }\label{fig:CompareDecay}
\end{figure}

In the GPE results, no such decay is observed.  Figure \ref{fig:CompareDecay} shows a comparison between one experimental run and a GPE calculation for very similar parameters ($\delta = -0.1$~kHz).  Indeed, GPE calculations to 64 ms show no sign of damping. Besides the possibility of the damping arising from technical sources, it may be due to thermal or other stochastic effects not included in the $T=0$ mean field calculation.

\section{Role of trap anisotropy}

\begin{figure}
\begin{center}
\includegraphics[scale=0.4,clip=true]{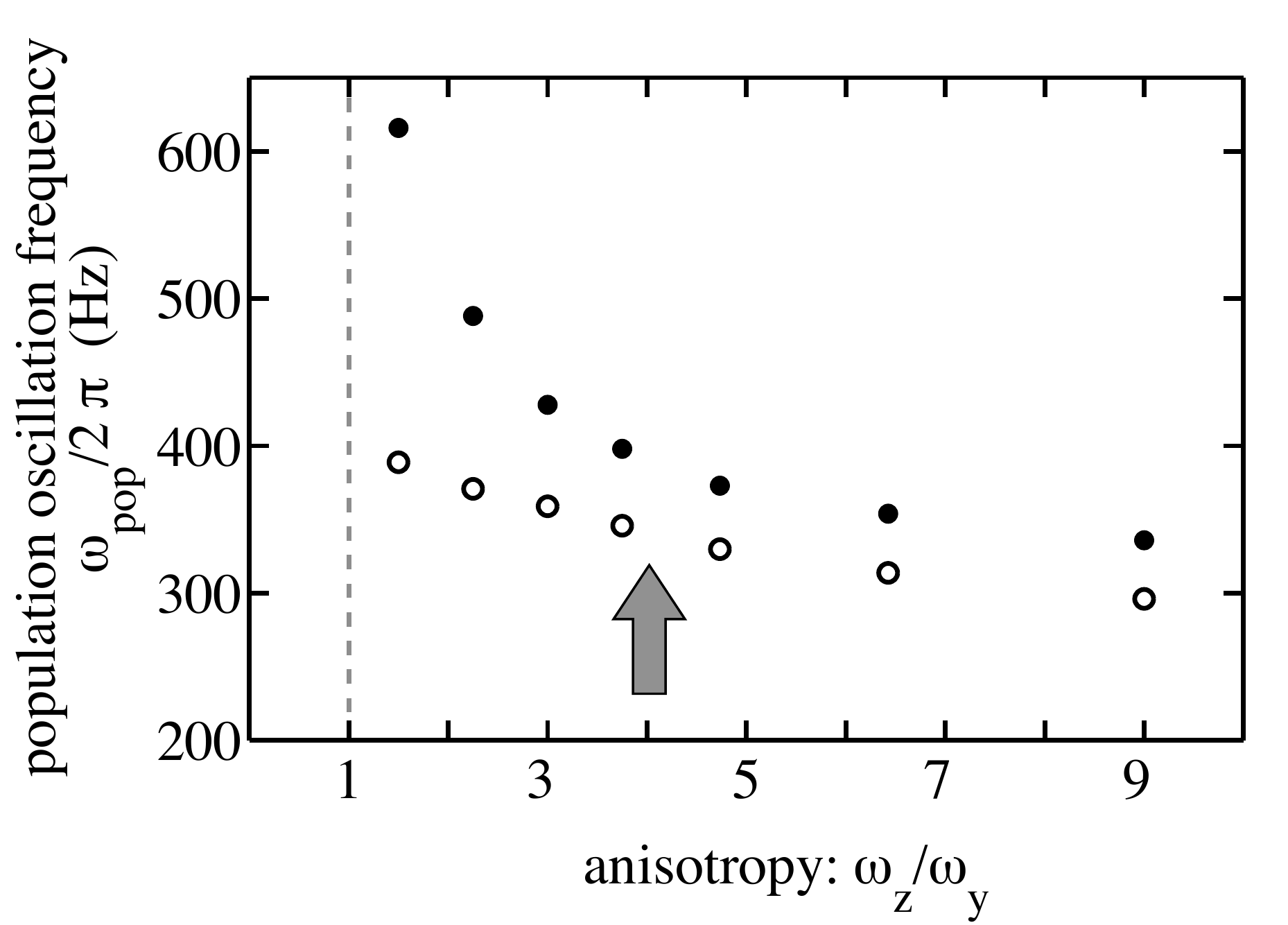}
\caption{Mode frequencies for $m = 0$ (open) and $m = 2$ (closed) as a function of trap anisotropy.  Grey arrow indicates the anisotropy used in this experiment. These calculations use the approximate potential Eq.~(\ref{eqpot}).  These  are calculated with fixed $\omega_z$ and decreasing $\omega_y$}\label{fig:anisotropy}
\end{center}
\end{figure}

We studied the role of the trap anisotropy (i.e., $\omega_z \ne \omega_y$) by observing the transformation of the $m= 0$ and $m = 2$ modes as the trap is deformed from axially symmetric to strongly axially
anisotropic, in presence of a purely anharmonic potential along $x$, using the simplified potential, Eq.~(\ref{eqpot}). The dipole
perturbation excites both modes as soon as the axial symmetry is broken,
and the spectrum shows a second frequency growing in strength as
the axial anisotropy is increased. The values of the mode frequencies as a
function of $\omega_z / \omega_y$ are shown in Fig.~\ref{fig:anisotropy}. Close to axial symmetry, the lower frequency depends only slightly on the transverse confinement, indicating that
the $m = 0$ mode is a dominant component of the Bogoliubov excitation. Motion is primarily along the splitting direction without oscillations in the transverse directions.

Sufficiently far from axial symmetry, both frequencies start to
decrease with increasing anisotropy and show a similar behavior. In particular, the experimental trapping conditions correspond to the point $\omega_z \approx 4\omega_y$, as indicated in Fig.~\ref{fig:anisotropy}, where the two modes begin to show a similar dependence on transverse
confinement. This strongly suggests that for such high axial anisotropy,
each Bogoliubov mode is mainly a combination of the two original $m = 0$
and $m = 2$ modes at axial symmetry.

\section{Full description of RWA potential}
The double-well potential is created through a coupling between static and rf magnetic fields.  In the dressed state picture, these combine to form the effective potential \cite{LesanovskyPRA2006}
\begin{equation}
U_\mathrm{RWA}(\mathbf{r}) = m^\prime_\mathrm{F}\sqrt{\left[ \hbar \omega_\mathrm{rf} - g_\mathrm{F} \mu_\mathrm{B} B_\mathrm{S}(\mathbf{r}) \right]^2 + \left[\frac{g_\mathrm{F} \mu_\mathrm{B} B_{\mathrm{rf,\perp}}(\mathbf{r}) }{2} \right]^2}	
\label{eq:Ueff}
\end{equation}
where $m^\prime_F$ is the adiabatic magnetic quantum number, $g_F$ is the Land\'e $g$-factor, $\mu_B$ is the Bohr magneton, $B_\mathrm{S}(\mathbf{r})$ is the static magnetic field, described by an Ioffe-Pritchard potential, and $B_\mathrm{rf,\perp}(\mathbf{r}) = |\mathbf{B}_\mathrm{S}(\mathbf{r}) \times \mathbf{B}_\mathrm{rf}(\mathbf{r})|/|\mathbf{B}_\mathrm{S}(\mathbf{r}) |$ is the component of the oscillating magnetic field locally perpendicular to the static field at each point, $\mathbf{r}$.

The static magnetic trap arises a result of the combination of current flowing through the `Z'-wire on the chip, an external bias field, and an external Ioffe field.  In combination, these create an Ioffe-Pritchard style trap, a static magnetic field $\mathbf{B_\mathrm{S}} = B_x \mathbf{\hat{x} }+  B_y  \mathbf{\hat{y}} +  B_z  \mathbf{\hat{z}}$, whose components are described by 
\begin{align}
&B_x(x,z) = B^\prime x - \frac{B^{\prime\prime}}{2}xy, \\
&B_y(x,y,z) = B_\mathrm{S}(\mathbf{0}) + \frac{B^{\prime\prime}}{2}\left(y^2 - \tfrac{1}{2}(x^2+z^2)\right),\\
\textrm{and}\\
&B_z(y,z) = -B^\prime z -\frac{B^{\prime\prime}}{2}yz.
\end{align}
In the limit of a small cloud, the static potential is well-approximated by a harmonic trap, characterized by radial and axial trapping frequencies $\omega_{x,z} $ and $\omega_y$. In terms of these measurable values, the static trap-bottom term, the gradient term, and the curvature term are given by
\begin{align}
B_\mathrm{S}(\mathbf{0}) &= \frac{2 \hbar \omega_\mathrm{TB}}{m^\prime_\mathrm{F} g_\mathrm{F} \mu_\mathrm{B}}\\
B^\prime &= \sqrt{\frac{m B_\mathrm{S}(\mathbf{0})}{m^\prime_\mathrm{F} g_\mathrm{F} \mu_\mathrm{B}}\left(\omega_{x,z}^2+\frac{\omega_y^2}{2}\right)}\\
B^{\prime\prime} &= \frac{m \omega_y^2}{m^\prime_\mathrm{F} g_\mathrm{F} \mu_\mathrm{B}},
\label{eq:Bterms}
\end{align}
respectively, where we define $ \omega_\mathrm{TB} = \mu_\mathrm{B} g_\mathrm{F} B_\mathrm{S}(\mathbf{0})/\hbar$  as the ``trap bottom'' frequency.

Typical values for the parameters in Eqs.~(\ref{eq:Bterms}) and (\ref{eq:Ueff}) are:  $(\omega_{x,z},\omega_y) = 2 \pi \times (1310, 10)$~Hz, $\omega_{\mathrm{TB}} = 2\pi \times 787$~kHz, $B_\mathrm{rf,\perp} = 240$~mG, $\omega_{y,0} = 2\pi \times 95$~Hz, and in the $\ket{F = 2, m^\prime_\mathrm{F} = 2}$ state of \re we use, $m_\mathrm{F}^{\prime} g_\mathrm{F} = 1$.   

\section{Corrections to the rotating-wave approximation}

To calculate our trapping potential, Eq.~(\ref{eq:Ueff}) assumes the rotating-wave approximation (RWA), but as discussed in \cite{HofferberthPRA2007}, the RWA fails for large Rabi frequencies.  We study the effect of the beyond-RWA effects for our trap and find that we can account for the difference between the approximate and full potentials by simply shifting the detuning by a fixed amount.

\begin{figure}
\includegraphics[scale=0.43,clip=true,trim=0.0cm 0.0cm 0cm 0cm]{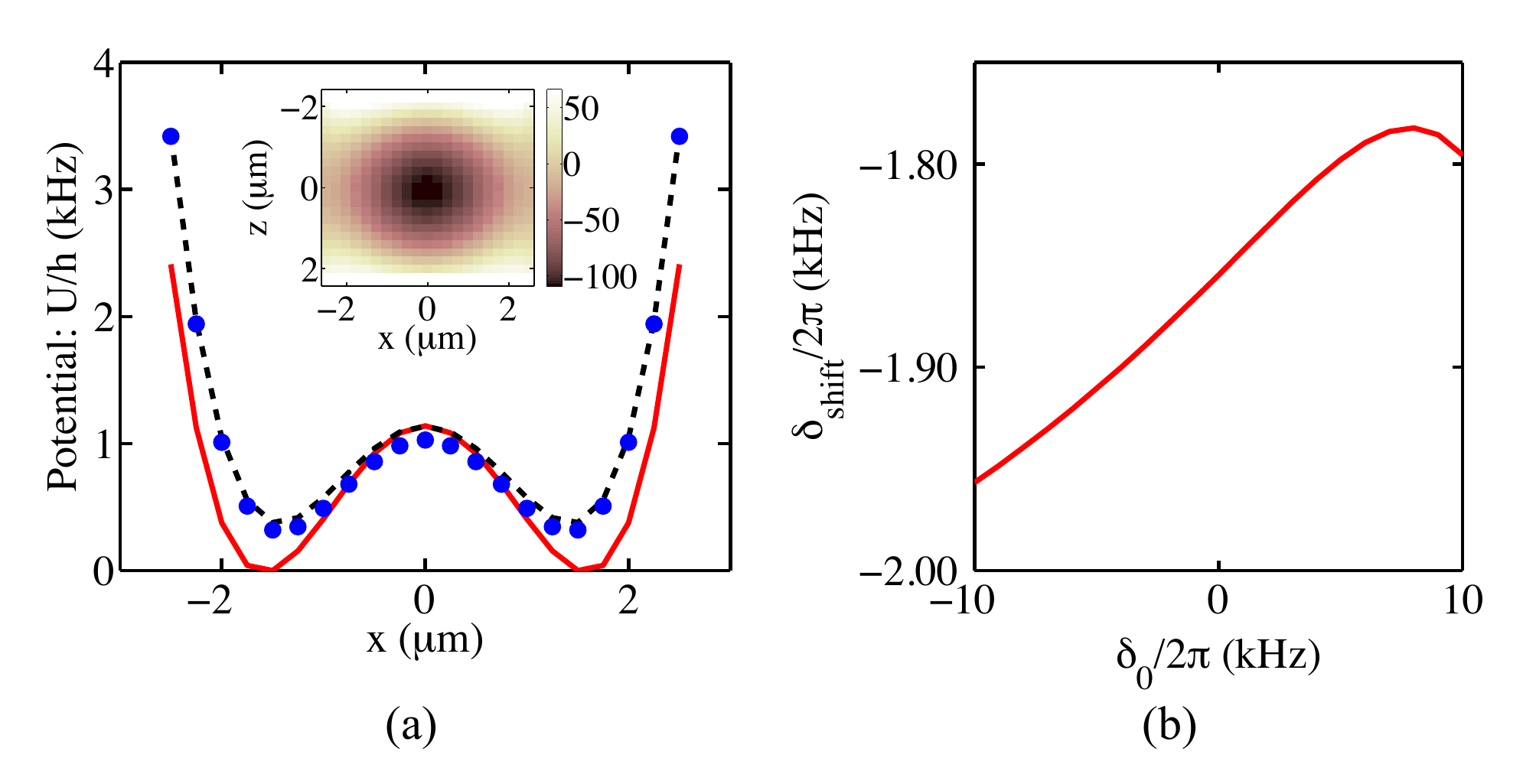}
\caption[Comparisons between RWA and full potential calculations.]{Comparisons between RWA and full potential calculations.  (a) Potential energy curve through $y = z = 0$ with $\delta_0/2\pi = 0$, calculated using full potential (blue dots), RWA approximation (red solid line), and RWA approximation with fitted shift  of $\delta_\mathrm{shift}/2\pi =  -1.85$~kHz (black dashed line).  Inset: difference between full potential and RWA potential (in Hz) with shift over entire 2D plane at $y = 0$ used for fit.  Color bar indicates in Hz the difference of potentials.  (b) Fitted detuning shift as a function of detuning, i.e., the number one should subtract from the detuning in the RWA expression to get the best estimate of the potential.}\label{fig:BeyondRWA}
\end{figure}

We calculate the full potential in a two-dimensional plane at $y = 0$ for our trap at a particular detuning, $\delta_0$, as described in \cite{HofferberthPRA2007}.  This 2D contour is fit to 
\begin{equation}
U_\mathrm{RWA} = m'_\mathrm{F}\mathrm{sgn}(g_\mathrm{F}) \hbar \sqrt{(\delta(\mathbf{r})-\delta_\mathrm{shift})^2 + \Omega^2},
\label{eq:Ueff}
\end{equation}
at $y=0$, which is just Eq.~(\ref{eq:Ueff}) without the compression term, where $\delta_\mathrm{shift}$ is the only fit parameter and describes a shift of the detuning.  We compare the shape of the full potential to the RWA potential with the shift and find that they are very similar.  The shift is calculated for all detunings used in this work and is roughly uniform for the range explored (Fig.~\ref{fig:BeyondRWA}).  We apply a shift $\delta_0 \rightarrow \delta_0 + 2 \pi \times 1.9$~kHz to each of the detunings used with the RWA in this work.  

The shift we find is of the opposite sign to that found in  Ref.~\cite{HofferberthPRA2007}.  Compared to the potential used in that work, our Rabi frequency is much smaller and our detuning much closer to zero, and we have confirmed that the shift changes sign for larger $\Omega$ and large negative detunings.    

\section{Systematic trap-bottom shifts}

As noted in the main text, we shift the data by a fixed detuning for display in Fig.~3.  Despite the discrepancy between the calculated and measured values of the detuning, the data matches the GPE in terms of the shape of the curves and the slope of the population oscillation frequency as a function of $V_b/\mu$.  For this reason, we are satisfied that the shift we are applying is acting to account for an unknown systematic uncertainty in the determination of the trap bottom value $B_\mathrm{S}(\mathbf{0})$. 

After taking into account all known systematic shifts, which include the beyond-RWA effect described above and calibrations in measuring the trap bottom $B_\mathrm{S}(\mathbf{0})$, we fit the experimental data to the GPE simulation data using a single-parameter least-squares fit, where the fitting parameter acts to slide the data along the detuning axis.  We find that a shift of  $2\pi \times (5.1 \pm 0.1)$~kHz accounts for the difference between the experiment and the GPE.  This shift is in the opposite direction to the beyond-RWA corrections.  Possible sources of this discrepancy include systematic errors in determining the static trap bottom $B_\mathrm{S}(\mathbf{0})$, or imperfections in the polarization of $\mathbf{B}_{\rm RF}$ due to the proximity of the fields to the chip and its copper support.

\section{Atom number}

To calibrate the atom number, we use standard absorption imaging to measure the thermal fraction of clouds above and below the condensate temperature.  We determine the total atom number by measuring the total absorption of the cloud, and the thermal number by fitting the wings to a Bose-Einstein distribution and integrating under the entire curve to extract thermal atom number.   The temperature of each condensate is determined by fitting the wings to a Gaussian.  

To find $T_C$, the condensation temperature, we plot the condensate fraction as a function of temperature.  We determine the temperature at which the condensate fraction is first non-zero, and find the number of atoms to which this corresponds.  Using the relationship between condensation temperature and atom number, including finite size and interaction effects [S1], we can determine the condensation temperature to $\pm$9\% ($T_C = 640 \pm 40$~nK).  Propagating this error through to atom number, we arrive at a calibration factor $N_\mathrm{actual} = N_\mathrm{measured}\times (1.3 \pm 0.3)$, which accounts for the systematic uncertainty in our atom number, $N = 6600 \pm 1700$.

The number, $N = 8000$ was chosen for the calculations  because this is the number within the systematic uncertainty for which the best agreement is found for mode amplitudes (Fig. 4).  The same $N = 8000$ is used in the Josephson model and hydrodynamic approximations.

\section{Data analysis}

To analyze the time series data, as in Fig.~2(a), we use a Fourier transform (FT).  To prepare the data, we eliminate the offset from $\mathcal{Z}  = 0$ components by subtracting from each point the mean, where the mean might be non-zero due to a small equilibrium imbalance in the system.  We smooth the transformed data by padding the time series with zeros to a total of 1024 points.  

When extracting the peak locations from the FT, we eliminate the points below the frequency given by $1/t_{\mathrm{tot}}$, where $t_{\mathrm{tot}}$ is the longest hold time.  We then find the two peaks with the maximum height and use these are our data points in Fig.~3.  We plot the amplitudes-squared of the FTs in the color-map in Fig.~3 behind the data.  The data are linearly interpolated numerically between the values of $V_\mathrm{b}/\mu$ ($\delta_0$) at which the data were measured.  

The uncertainty in the frequency measurement is found by simulating data with the same level of noise as the original time series.  The quantity of noise is determined by fitting the time series to a 2-frequency decaying exponential function
\begin{equation}
\mathcal{Z}(t) = e^{-t/\tau} \left[a_1 \sin(2\pi \nu_1(t-t_{01})) + a_2 \sin(2\pi \nu_2(t-t_{02}))\right],
\label{eq:2freq}
\end{equation}
where $\tau$ is a time constant for decay, $a_{1(2)}$ is the amplitude of the first (second) frequency component, and $\nu_{1(2)}$ is the first (second) frequency component, and $t_{01(02)}$ is the constant accounting for the phase shift of the first (second) component.  The standard deviation of the residuals from this fit gives the noise level.  We simulate data 100 times with the same parameters as those given by the fit, with the same total time and density of points, but with different randomized instances of Gaussian noise whose standard deviation is the same as that measured.  Taking the frequency measurements from each of these trials, we determine the smallest range inside of which 68\% of the measurements lie.  This confidence interval is used as the uncertainty in the frequency measurement.

The noise floor in the FT is established in a similar fashion.  Using the result for the noise level from the time series, we simulate pure Gaussian noise and take the FT of this.  The noise floor we show is the mean plus one standard deviation of the maximum peak amplitudes found in 100 such simulations.

The amplitudes used in determining the ratio $R_1$, shown in Fig.~4, are given by the values determined by the fit (Eq.~(\ref{eq:2freq})).  The uncertainties in these values are determined in a similar way to those in the frequencies; we use the noise level in the residuals of the fit, simulate and fit 100 sets of data with similar parameters, and use the 68\% confidence interval of these results to represent our uncertainty.  

One significant difference between the calculated and measured quantities is that the calculated amplitudes display no decay.  The measured values, which come from the fits to the Eq.~(\ref{eq:2freq}), rely upon the fitting routine to extrapolate backwards in time to deterimine the $t = 0$ amplitudes.  The uncertainty associated with this process results in the scatter in the measurements, and may be a cause of some of the discrepancy between the calculated and measured values.

\bibliographystyle{prsty}

%\bibliography{/Users/Lindsay/Documents/Thesis/TeX/LJLbibThesis}

\begin{thebibliography}{99}
\bibitem{footnoteZ}
{When the average value of $\mathcal{Z}$ differs from zero, we subtract the average $\bar{\mathcal{Z}}$ from all values of $\mathcal{Z}(t)$.  In all experiments,  $|\bar{\mathcal{Z}}| < 0.05$.}
\bibitem{footnoteTMM}
{We checked this numerically by deforming our trap into a fully harmonic axially symmetric trap, and following the mode frequencies throughout this process.}
\end{thebibliography}

{\small 
\noindent[S1]
S. Giorgini, L.~P. Pitaevskii, and S. Stringari, Phys. Rev. A {\bf 54},  R4633
  (1996).

\noindent[S2]
A. Smerzi and S. Fantoni, Phys. Rev. Lett {\bf 78},  3589  (1997).
}

\end{document}